\documentclass[aps,prl,preprint,showpacs,groupedaddress]{revtex4}
\usepackage{epsfig}
\usepackage{bm}

\begin{document}

\title
{An absolutely secure QKD scheme with no detection noise, entanglement and classical communication\footnote{The material presented in this paper is described in a CUNY-TAMU patent disclosure \cite{8}}} 
\author{J\'{a}nos A. Bergou}
\affiliation{Department of Physics, Hunter College, City University of New York, 695 Park Avenue, New York, NY 10021, USA}
\author{Laszlo B. Kish}
\affiliation{Department of Electrical Engineering, Texas A \& M University, College Station, TX 77843-3128, USA}
\date{\today}
\begin{abstract}
A new quantum communication scheme is introduced which is the quantum realization of the classical Kish-Sethuraman (KS) cipher. First the message is bounced back with additional encryption by the Receiver and the original encryption is removed and the message is resent by the Sender. The mechanical analogy of this operation is using two padlocks; one by the Sender and one by the Receiver. We show that the rotation of the polarization is an operator which satisfies the conditions required for the KS encryption operators provided single photons are communicated. The new method is not only simple but has several advantages. The Evasdropper extracts zero information even if she executes a quantum measurement on the state. The communication can be done by two publicly agreed orthogonal states. Therefore, there is no inherent detection noise. No classical channel and no entangled states are required for the communication.
\end{abstract}
\maketitle

\section{Introduction}
\begin{figure}[ht]
\epsfig{file=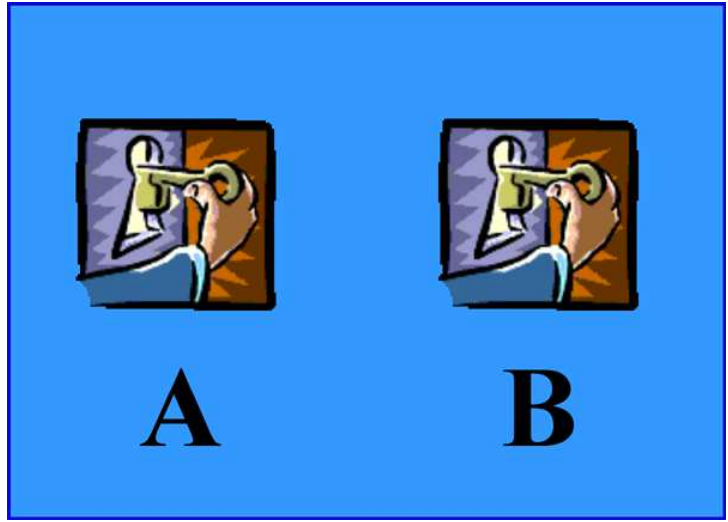, height=7cm}
\caption{Mechanical analogy of the Kish-Sethuraman cipher. The operations represent locking ($A$ and $B$) and opening ($A^{-1}$ and $B^{-1}$) and they are to be carried out by the Sender ($A$ and $A^{-1}$) and Receiver ($B$ and $B^{-1}$), respectively.}
\label{Fig1}
\end{figure}

Recently, Kish and Sethuraman proposed a new method \cite{1}, which is claimed to be an absolute secure data encryption, by using only classical information. Sethuraman \cite{2} has identified a mechanical analogy of the method with a mail-carrier-box and two-padlocks, see Fig. \ref{Fig1}. The Sender is using a padlock to lock the box (operation $A$) before the mail is sent. When the Receiver receives the box, he is using another padlock to lock the box again (operation $B$). Then the double-locked box is sent back to the Sender where the first padlock is removed (operation $A^{-1}$). Then the Sender sends the box again, which is still locked by the Receiver's padlock, back to the Receiver where the box is opened (operation $B^{-1}$) thus the mail is delivered. In conclusion, at the Kish-Sethuraman (KS) chiper (so named by Klappenecker \cite{3,4}), the following operations are carried out on the $t^{th}$ bit (or word, etc.), m(t) of the message:
\begin{eqnarray}
m(t) &\stackrel{S}{\Longrightarrow}&  A(t) m(t) \stackrel{R}{\Longrightarrow} B(t) A(t) m(t) \nonumber \\
&\stackrel{S}{\Longrightarrow}& A^{-1}(t) B(t) A(t) m(t) = B(t) m(t) \nonumber \\
&\stackrel{R}{\Longrightarrow}& B^{-1}(t) B(t) m(t) = m(t) \ ,
\label{eq1}
\end{eqnarray}
where the arrows with S and R mean that the resulting signal was generated by the Sender and the Receiver, respectively.

At the third step, the presumed mathematical condition is:
\begin{equation}
A^{-1}(t) B(t) A(t) m(t) = B(t) m(t) \ ,
\label{eq2}
\end{equation}
which was identified by Bergou \cite{5} as commutativity of the operators:
\begin{equation}
A(t) B(t) = B(t) A(t) \ .
\label{eq3}
\end{equation}
The commutativity is necessary to get through the message to the Receiver.

Another mathematical condition concerns the security of the KS cipher: \newline 
{\underline{Condition 1.}} {\emph{It should be impossible to determine the message and the actual operator values by recording and comparing the publicly sent signals $A(t)m(t)$, $B(t)A(t)m(t)$, and $B(t)m(t)$.}}

This condition simply ensures that a comparison of the data sequences before and after the application of the  operators will not provide enough information to identify the data or the actual operator code.

A. Klappenecker \cite{3} pointed out that operators in the RSA public key cryptosystem satisfy Eq.\ (\ref{eq3}) and Condition 1. Further discussions with A. Singer \cite{6,7} made it obvious that, for the absolute security, one more condition is required which has been implicitly assumed but not explicitly stated: \newline
{\underline{Condition 2.}} {\emph{The Sender and Receiver share no information and submit no public information about the actual operators A(t) and B(t) (except their class which is public information).}}

Indeed, the RSA system is not absolutely secure because of the shared public information about the operators. Thus the system can be broken if the Eavesdropper has extraordinary calculational power for factoring (is, for example, in the possession of a quantum computer). 

In conclusion, with Conditions 1 and 2 the KS cipher (or any cipher, for that matter) would be absolutely secure, and Eq.\ (\ref{eq3}) is needed for the functioning of this crypto system. However, it seems to be extremely difficult to find classical operators which satisfy all three of these requirements. For example, the classical KS cipher does not meet Condition 1. The Eavesdropper can record the three sequences $A(t)m(t)$, $B(t)A(t)m(t)$, and $B(t)m(t)$ with no difficulty. Then simply by comparing the first and the second sequence, she can read out $B(t)$ and applying the inverse, $B^{-1}$, to the third sequence she will have full access to the key. Similarly, the RSA operators satisfy Eq.\ (\ref{eq3}) and Condition 1 but not Condition 2. 

\section{The solution: Quantum realization}

In the following we will introduce a straightforward quantum generalization of the above two-padlock method and show that the quantum version provides absolute security for QKD. In particular, as we shall demonstrate, all of the above requirements, Eq. (\ref{eq3}) and Conditions 1 and 2 are met.

First we replace the individual classical bits in the message by qubits, i.e.\ we communicate the single bits by single quantum systems with two states, such as polarization or spin or two-level atoms, etc. For the sake of a concrete example, but by no means restricting generality, let us suppose that we communicate with single photons and the information-bit is represented by two orthogonal polarization states, $m(0^{o})$ for horizontal and  $m(90^{o})$ for vertical polarization, respectively. The quantum realization of Eq. (1) and that of the two padlock arrangement of Fig. 1 is shown in Fig. 2. Alice, the Sender, generates a qubit randomly in either the state 0 or the 1.  Then, in the next step she generates a random unitary operator, $A$, which, in this case is a polarization rotation by a random amount and applies it to the qubit she generated. She then sends this qubit which has now a completely random orientation over to Bob, the Receiver. Bob generates another random unitary operator, $B$, which rotates the polarization by an additional random amount and applies it to the qubit he just received and then sends it back to Alice. Alice, at her end, applies $A^{-1}$ to the qubit she receives and redirects it to Bob. In the final step Bob applies $B^{-1}$ to the qubit and then measures the qubit in the orthogonal basis 0 and 1. Whichever detector clicks will tell the state of the initially prepared photon uniquely.

It is easy to see that these steps exactly correspond to the ones in Eq.\ (\ref{eq1}). Furthermore, the operators $A$ and $B$ are polarization rotations, which are randomly generated at the communication of each bits, and therefore they obviously satisfy Eq. (\ref{eq3}). Condition 1 is also clearly satisfied; at each step of the communication scheme the sates that travel between Alice and Bob are completely random and unknown quantum states can not be determined from a single measurement. Condition 2 is also trivially met; the keys, $A^{-1}$ and $B^{-1}$, never leave Alice's and Bob's site, respectively. The mirrors $M_{S}$ at the Sender's site and $M_{R}$ at the Receiver's site have the only role of redirecting the photons and can be replaced by optical fibers or similar devices. The polarization beam splitter PBS directs the photons to the relevant detector. The resulting quantum communicaton scheme has superior properties. It is very simple, provides absolutely no  information to the Eavesdropper, has no inherent detection noise like other quantum communicators, and does not need rely on the use of a classical channel or entangled states. 

\begin{figure}[ht]
\epsfig{file=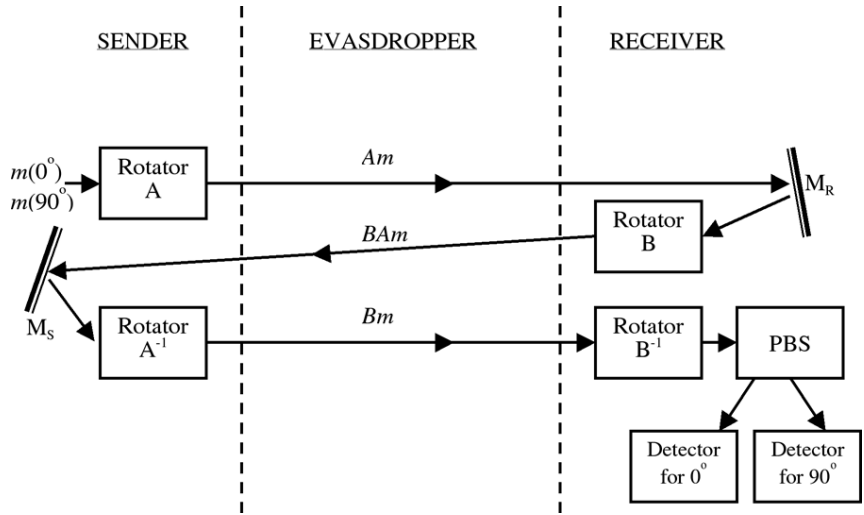, height=7cm}
\caption{Quantum realization of Eq. (1) and the corresponding two-padlock system in Fig. 1. The polarization rotations $A$ and $B$ are randomly generated at communicating each bits. The mirrors $M_{S}$ and $M_{R}$ at the Sender's site and Receiver's site, respectively, have the only role of redirecting the photons and can be replaced by optical fibers or similar devices. The polarization beam splitter PBS directs the photons to the relevant detector.}
\label{Fig2}
\end{figure} 

\section{Discussion and summary}

The new method has several advantages over exisiting proposals and we will briefly list them here, along with a short discussion of each of these features. 

1. The photons seen by the Eavesdropper have random polarizations. Moreover, from a quantum measurement on a single photon the Evasdropper is unable to identify the polarization. Thus the  absolute security conditions 1 and 2 are both satisfied and the Eavesdropper has zero information even if she measures the photons. This is in sharp contrast to previously proposed schemes where the Eavesdropper could get a finite amount of information but her presence could be detected.

2. The ultimate encoding of the key is into polarization eigenstates. Thus the inherent quantum detection noise originating from the necessity of detecting non-orthogonal polarization states in other quantum communication systems does not exist here.

3. Another feature is that the presence of an Eavesdropper can be found out in a simple manner.  Although it should be noted that in this scheme the Eavesdropper has no information to gain, her ultimately unsuccessful activities can still be detected. For example, if the Eavesdropper is using a photon amplifier to gain information, she introduces a quantum detection (cloning, amplification) noise in the system, which can immediately be detected, provided each bit is sent at least two times and the results are then publicly compared. Let us consider quantum cloning as an example. Since perfect cloning of a quantum state is impossible Eve has to rely on the next best thing which is  approximate cloning. In each step, the Eavesdropper can reach a maximum fidelity of 5/6 using the optimal universal quantum cloning machine. Since there are three exchanges between Alice and Bob this means that Eve can reach a maximum fidelity of $(5/6)^{3}=0.53$ which is hardly better than a completely random guess and is, thus, useless for any security breach. In fact, even this small window of opprtunity can be removed by using an obvious four-padlock extension of the above scheme (Alice and Bob each applies two quantum padlocks initially, and in each subsequent exchange they remove one of them, requiring five exchanges altogether).

4. Since Eve can, in principle, gain no information, no classical communication channel is needed. Single photons can be sent at each clock period in a synchronized way. 

5. The scheme is not based on entangled photon states.

In summary, the above scheme provides absolute security for secret communication, QKD in particular, between Alice and Bob since the communication is via pure noise from the point of view of an Eavesdropper. In no step of the scheme is there a possibility for Eve to access useful information and, thus, the need for classical communication is also eliminated. The proposed scheme can be generalized in many different ways and opens up several novel possibilities for absolutely secure  communication.

Note added: Recently we became aware that a very similar idea has actually been published already \cite{kye} but the order of adding and removing the quantum padlocks is different and therefore we feel that our method provides more security.

\begin{acknowledgments}
This research was partially supported by a grant from the Humboldt Foundation (JB) and by a Grant from PSC-CUNY. JB also acknowledges helpful discussions with Prof. W. Schleich and his group during a visit to the University of Ulm. The material presented in this paper is described in a CUNY-TAMU patent disclosure \cite{8}.
\end{acknowledgments}

\bibliographystyle{unsrt}

\begin{thebibliography}{99}

\bibitem{1}L.B. Kish and S. Sethuraman, Non-breakable data encryption with classical information, Fluctuation and Noise Letters {\bf 4}(2), C1-C5 (2004).

\bibitem{2}Swaminathan Sethuraman, personal communication (June 2004).

\bibitem{3}A. Klappenecker, Remark on a "Non-breakable data encryption" scheme by Kish and Sethuraman, Fluctuation and Noise Letters {\bf 4}(4), C25-C26 (2004).

\bibitem{4}L.B. Kish, Response to Klappenecker's Remark on a Non-Breakable Data Encryption Scheme by Kish and Sethuraman, Fluctuation and Noise Letters {\bf 4}(4), C27-C29 (2004).

\bibitem{5}J.A. Bergou, personal communication (April 2004).

\bibitem{6}A. Singer, personal communication at the UPoN'04 conference, June 2005, Gallipolli, Italy.

\bibitem{7}L.B. Kish, S. Sethuraman and P. Heszler, Non Breakable Data Encryption with Classical Information? Proc. Fourth International Conference on Unsolved Problems of Noise, Lecce, Italy, June 6-10, 2005, American Institute of Physics Press (2005), in press.

\bibitem{8}J.A. Bergou and L.B. Kish, A New Quantum Communicator with Enhanced Security, no Detection Noise, no Entanglement and no Classical Channel, CUNY-TAMU patent disclosure, July 5, 2005.

\bibitem{kye}W.-H. Kye {\it et al.}, \prl {\bf 95}, 040501 (2005).
\end{thebibliography}

\end{document}